\documentclass[floatfix,onecolumn,pre]{revtex4}
\usepackage[pdftex]{graphicx}
\usepackage{amssymb}
\usepackage{amsmath}
\usepackage{latexsym}
\usepackage{bm}
\usepackage{times,subfigure}
\begin{document}

\title{Modelling unidirectional liquid spreading on slanted microposts}
\author{Andrea Cavalli}
\affiliation{Department of Micro- and Nanotechnology, Technical University of Denmark, DTU Nanotech, Building 345 East, DK-2800, Kongens Lyngby, Denmark. E-mail: acav@nanotech.dtu.dk}
\author{Matthew L. Blow}
\affiliation{Centro de F\'{i}sica Te\'{o}rica e Computacional, Instituto de Investiga\c{c}\~{a}o Interdisciplinar, Av. Prof. Gama Pinto, 2, P-1649-003 Lisboa, Portugal.}
\author{Julia M. Yeomans}
\affiliation{The Rudolf Peierls Centre for Theoretical Physics, Oxford University, 1 Keble Road, Oxford OX1 3NP, England}
\date{\today}

\begin{abstract}
A lattice Boltzmann algorithm is used to simulate the slow spreading of drops on a surface patterned with slanted micro-posts. Gibb's pinning of the interface on the sides or top of the posts leads to unidirectional spreading over a wide range of contact angles and inclination angles of the posts. Regimes for spreading in no, one or two directions are identified, and shown to agree well with a two-dimensional theory proposed in Chu, Xiao and Wang \cite{chu2010uni}. A more detailed numerical analysis of the contact line shapes allows us to understand deviations from the two dimensional model, and to identify the shapes of the pinned interfaces.
\end{abstract}

\maketitle

\section{Introduction}
\begin{figure}
  \centering
  \includegraphics[width=80mm]{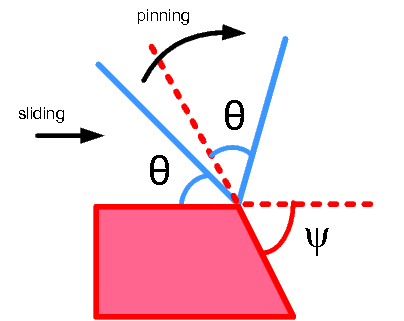}
  \caption{Gibbs' pinning on the corner of a post. The interface (blue line) remains pinned to the post over the range of angles indicated by $\psi$ as there is a free energy barrier to its moving in either direction.} 
  \label{fig:Gibbs}
\end{figure}

Unusual wetting and spreading properties of drops on natural and artificial surfaces can often be explained by the micro-structure of the substrate\cite{de2004capillarity}. For example drops spreading on superhydrophobic surfaces patterned with micron-scale ridges reach an elongated final state \cite{chen2005anisotropy, zhang2007anisotropic}, and a drop imbibing into a surface patterned with posts can form a faceted final configuration, which reflects the symmetry of both the lattice and of the posts themselves \cite{blow2011anisotropic}.\\

The physics behind this behaviour was first described by Gibbs, who pointed out that an interface can pin on the edge of a post over a range of angles, as illustrated in Fig.~\ref{fig:Gibbs}. The pinning occurs because there is a free energy penalty to the interface moving away from the edge in either direction as it would then have to form an angle with the adjacent surface which differs from the equilibrium contact angle. A pinning strength that depends on the lattice direction leads to the drop having one or more preferred directions of motion, and hence anisotropic drop movement and shapes.\\

In this paper we focus on unidirectional drop motion: where the symmetry of the underlying surface structure can pick out one easy direction of spreading \cite{hancock2012directional,extrand2007retention}. Such surfaces occur naturally, for example,  the unidirectional motion of droplets on butterfly wings  results from their ratchet-like structure  \cite{zheng2007directional,kusumaatmaja2009anisotropic}, and rye-grass leaves shed water in a preferred direction  \cite{guo2012directional} due to the asymmetric contact angle hysteresis. Microfabricated surfaces that  lead to uni-directional motion are a very recent development. Unidirectional spreading has been observed on bent silicon micro pillars \cite{chu2010uni}, while other authors \cite{malvadkar2010engineered,sekeroglu2011transport} were able to transport droplets on vibrating ratchet structures. Similar results where obtained using PDMS replicas of the naturally occurring asymmetric micro-texture \cite{guo2012directional} taken from rye grass, and anisotropic hysteresis was observed on printed ratchetlike surfaces \cite{barahman2011ratchetlike}. Despite the different material and geometries employed, the unidirectional liquid motion consistently reflects the asymmetry of the substrate on microscopic length scales. This highlights the importance of understanding the underlying physical phenomena involved.\\

In this paper, we use a two-phase lattice Boltzmann algorithm to model imbibition on an hydrophilic surface patterned with slanting posts, varying the contact angle of the substrate and the tilt angle of the posts. For a range of contact angles we observe a single, preferred spreading direction as observed in recent experiments  \cite{chu2010uni}. The results are in good qualitative agreement with a two-dimensional model of the uniaxial spreading proposed in \cite{chu2010uni}, and enable us to describe the corrections to the model for a three dimensional geometry.  By visualizing the shape of the contact line we describe in detail the mechanisms through which the interface pinning and de-pinning occurs.

\section{The model liquid}

\subsection{Governing equations}

To model a two phase system interacting with a surface, we apply a diffuse interface scheme. The thermodynamic state of the fluid is described at every point $\vec x$ and time $t$ by its density $\rho (\vec x, t)$. The free energy of the system, $\Psi$, is taken as a Landau double-well potential with the addition of a derivative term representing the surface tension, and a surface contribution of the form proposed by Cahn \cite{cahn1977critical}:
\begin{equation}
\label{eq:freeEnergy}
\Psi = \iiint\limits_D \left( \psi_b(\rho)- \mu_b \rho + \frac{1}{2}\kappa |\nabla\rho | ^2 \right)\; dV -\iint\limits_{\partial D}  \mu_s \rho \; dS.
\end{equation}
The first term in the integrand of (\ref{eq:freeEnergy}) is the bulk free energy density \cite{briant2004lattice}
\begin{equation}
\label{eq:bulkFreeEnergy}
\psi_b(\rho) = p_c \left[ \left( \nu^2-\beta \tau_W \right)^2- \left( 1-\beta \tau_W \right)^2  \right].
\end{equation}
where $\rho_c$ , $p_c$ , $\tau_W$ and $\beta$ are, respectively, the critical density, critical pressure, reduced temperature and a free parameter controlling the density difference between phases and $\nu= \frac{\rho-\rho_c}{\rho_c}$ is a normalised density. This potential leads to two equilibrium bulk densities $\rho_e = \rho_c \left( 1 \pm \sqrt{\beta \tau_W} \right) $. $\mu_b$  is a Lagrange multiplier constraining the total mass of fluid, while the third term in the free energy expression is an interface energy cost, tunable through the parameter $\kappa$, associated with density gradients. It allows for solutions with a diffuse interface between phases, with surface tension $\gamma$ and width $\chi $: 
\begin{equation}
\label{eq:interface}
\gamma = \frac{4}{3} \rho_c\sqrt{2 \kappa p_c (\beta \tau_W)^3}, \; \chi= \frac{1}{2}\rho_c \sqrt{\frac{\kappa}{\beta \tau_W p_c}}.
\end{equation}
The final term in Eq.~(\ref{eq:freeEnergy}) is the surface contribution to the free energy $\Psi$. When $\Psi$ is minimised this gives the boundary condition $ \partial_{\perp} \rho = - \mu_s/\kappa$ which fixes the value of the density at the solid surface. The Young contact angle $\theta$ at the surface is related to the surface chemical potential by \cite{briant2004lattice} 

\begin{eqnarray}
\label{eq:angle}
\mu_s = 2 \beta \tau_W \sqrt{2 p_c \kappa} \, \mathrm{sign} \left( \frac{\pi}{2} - \theta \right)\sqrt{\cos \frac{\alpha}{3}\left( 1- \cos \frac{\alpha}{3} \right)}, \; && \nonumber\\
 \alpha =  \mathrm{arccos}(\sin^2 \theta).&&
\end{eqnarray}
In our simulations, the main parameters are set as follows: $\kappa=0.01$, $p_c = 0.125$, $\beta = 1$, $\tau_W = 0.3$, $\rho_c=3.5$. The corresponding surface tension is $\gamma = 0.04$ and the surface thickness is $ \chi = 0.9$ (in simulation units).\\

The hydrodynamics of the fluid is described by the continuity and Navier-Stokes equations: 
\begin{equation}
\partial_{t} \rho+ \partial_{\alpha}(\rho u_{\alpha})
= 0 ,
\label{eq:continuity}
\end{equation}
\begin{eqnarray}
\nonumber
&& \partial_{t}(\rho u_{\alpha}) + \partial_{\beta}(\rho u_{\alpha}u_{\beta}) = \\ 
&&-\partial_{\beta}P_{\alpha \beta}+ \partial_{\beta} \left( \rho \eta \left[ \partial_{\beta} u _{\alpha} + \partial_{\alpha} u _{\beta} \right] + \rho \lambda \delta_{\alpha \beta } \partial_{\gamma} u _{\gamma} \right) , 
\label{eq:governing}
\end{eqnarray}
where $\mathbf{u} $ is the fluid velocity field and $\eta$ and $\lambda$ are the shear and bulk kinematic viscosities, respectively. $\partial_t$ represents a time derivative and $\partial_{\alpha}$, $\partial_{\beta}$ spatial derivatives (Einstein summation convention is assumed). The connection between the thermodynamic and fluid dynamic of the system arises through the pressure tensor $\mathbf{P} $, which is derived from the free energy (\ref{eq:freeEnergy}). Equations (\ref{eq:continuity}) and (\ref{eq:governing}) are solved using the Lattice Boltzmann method. Details of the implementation can be found in \cite{kusumaatmaja2009anisotropic,briant2004lattice}.

\subsection{Geometry}

\begin{figure}
  \centering
  \includegraphics[width=180mm]{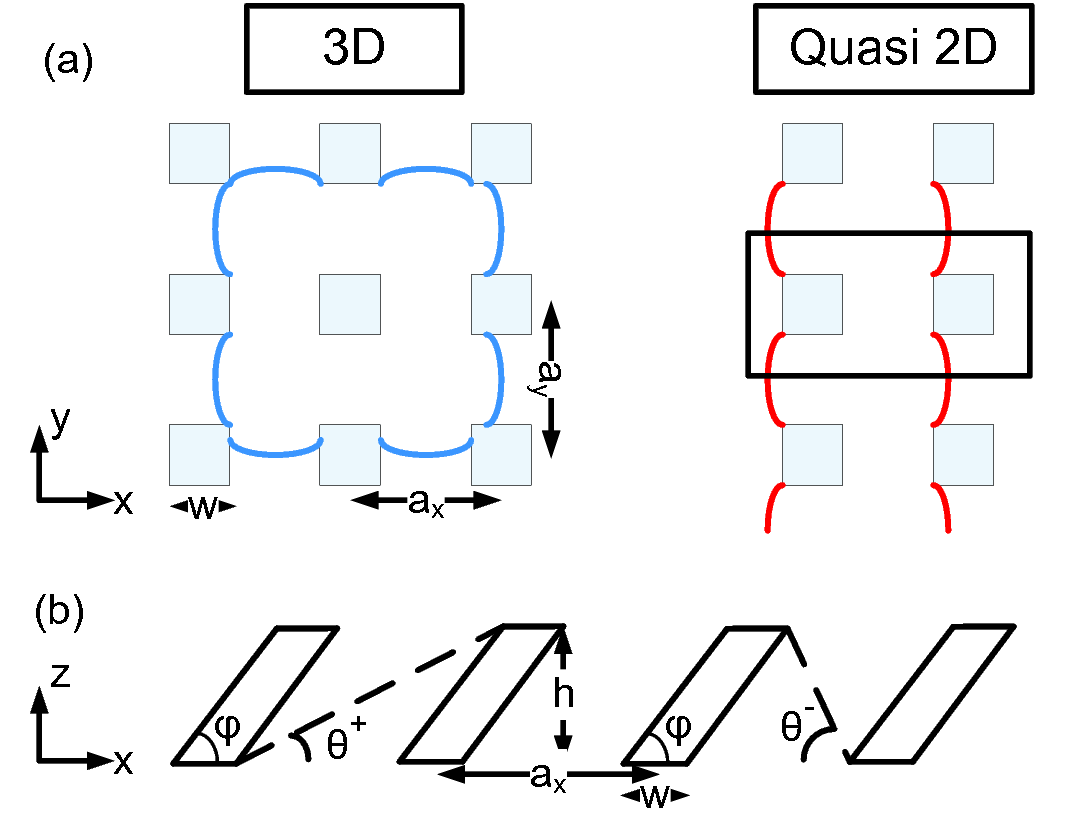}
  \caption{(a) Diagrams to contrast the quasi-2D and full 3D geometries used in the simulations. The red and blue lines represent typical interface positions. (b) Side view of the posts showing the interface geometry corresponding to the threshold for movement which is assumed in deriving Eq.~(\ref{eq:crit+}).} 
  \label{fig:layout}
\end{figure}

Each post has a square cross section of dimensions  $w$, which was typically chosen to be equal to 5 or 10 computational grid spacings. The posts are tilted at an angle $\phi$ to the positive $x$-axis, and extend to a height $h=4w$ above the surface. We start considering a rectangular array of posts with lattice constant $a_x=4.6w$ in the x direction and $a_y=4w$ in the y direction as shown in Fig.~\ref{fig:layout}. \\

Placing a sufficiently large drop on the surface to allow significant spreading through the posts is computationally expensive. Therefore we define a reservoir of fluid in the centre of the post array typically extending across $\simeq 10w$ and reaching to the top of the posts. If the fluid density inside the reservoir decreases below the equilibrium density of the liquid phase, new mass is slowly added to feed the imbibition. The contact angle of the liquid with both the posts and the substrate is $\theta$ which we vary in the range $30^o$ to $70^o$. This range is representative of different hydrophilic material, for example the polymers considered in\cite{chu2010uni} or \cite{vrancken2013anisotropic}.\\

We first consider a quasi-2D geometry which allows us to concentrate on the directional spreading in the $x$-direction. We choose a simulation box of length $40w$ in the $x$-direction, $4w$ in the $y$-direction and $6w$ along $z$, with periodic boundary conditions along both $x$ and $y$.  The reservoir spans the simulation box in $y$ corresponding to simulating a cylindrical drop with interfaces which lie, on average, parallel to the $y$-direction. The average of any fluid motion is along $x$. We then present results for a full 3D geometry, using a simulation box with typical dimensions $40w \times 32w \times 6w $, with periodic boundary conditions along $x$ and $y$. A schematic comparison of quasi-2D and 3D spreading is given in Fig.~\ref{fig:layout}a.
\section{Results}
\begin{figure}
\centering
\includegraphics[width=180mm]{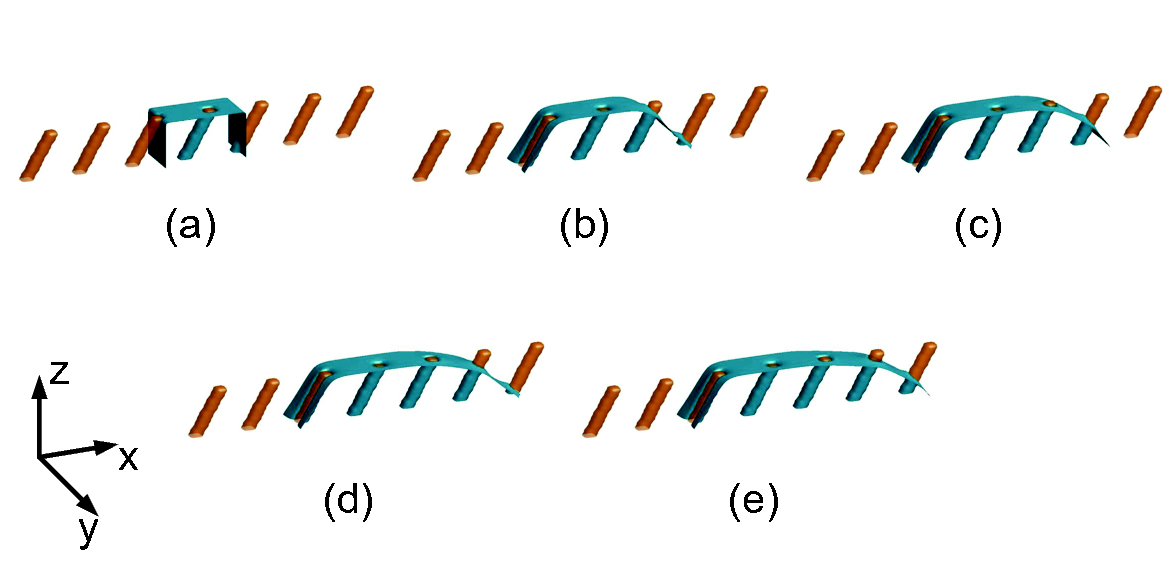}
\caption{Advancing front in the unidirectional spreading regime for contact angle $\theta = 45^o$ and post angle $\phi= 60^o$. This is a quasi-2D geometry with periodic boundary conditions along $y$. The snapshots (a)-(e) correspond to 0,1,2,4,5 $\times 10^4$ time steps.}
 \label{fig:1D}
\end{figure}
\begin{figure}
\centering
\includegraphics[width=180mm]{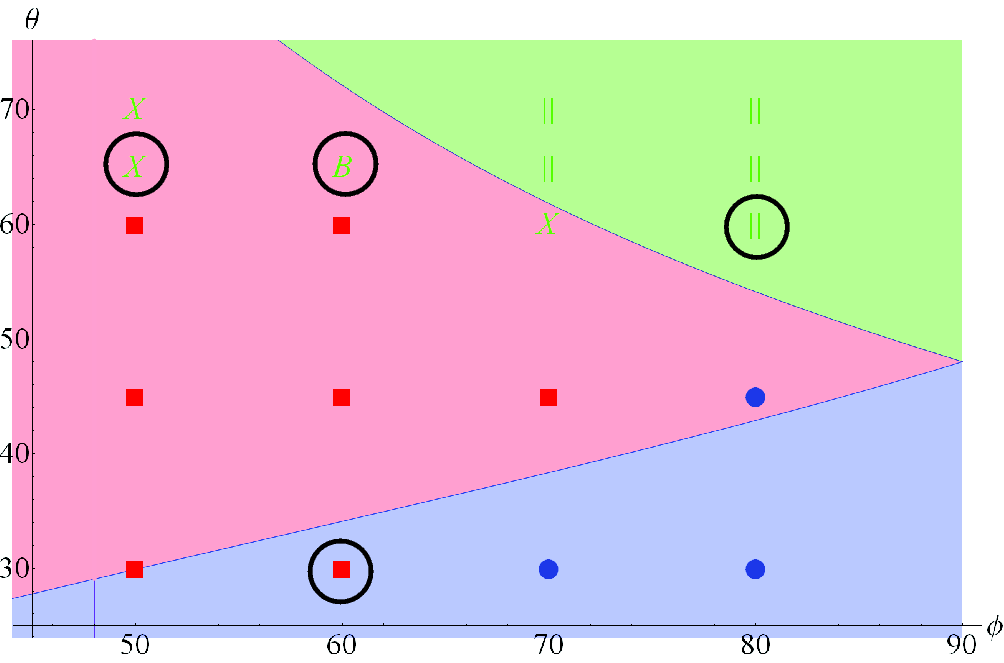}
\caption{Plot of the different wetting regimes in the quasi-2D geometry, as a function of wetting angle $\theta$ and post angle $\phi$. The other geometric parameters are: $a_x=4.6w $, $h=4w $. The legend is: blue/circles: bidirectional spreading, red/squares: unidirectional spreading, green: no spreading. The different green symbols represent different pinning modes in the forward direction, as described in Fig.~\ref{fig:modes}. Selected configurations (black circles) are shown in Fig.~\ref{fig:modes}. The background indicates wetting regimes from a theory assuming a 2D geometry, see Eq.~(\ref{eq:crit+}) \cite{chu2010uni}. }
 \label{fig:spreading}
\end{figure}
\begin{figure}
\centering
\includegraphics[width=180mm]{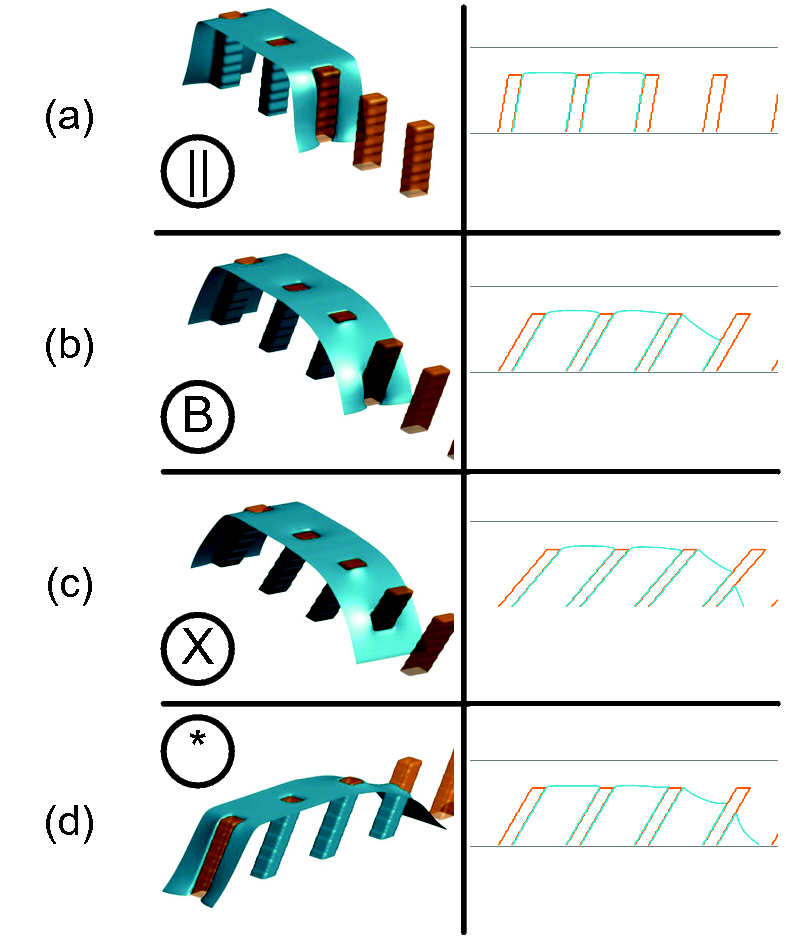}
\caption{ Different pinning configurations appearing in the quasi 2D simulations. The first three correspond to the easy spreading direction, $x$, while the last shows the typical pinning in the hard direction, $-x$. They can be identified with the circled points in Fig.~\ref{fig:spreading} by matching symbols. Cross sections taken through the centres of the posts are shown on the right as full blue lines. (a) If the post is almost vertical, the leading interface is disconnected and pinned to the vertical sides of the post ($||$ label). (b) For more pronounced post tilt the interface remains disconnected, but does not reach the top of the final post ($B$ label). (c) For large tilt the leading interface is connected and has reached the equilibrium contact angle on the substrate. However, in contrast to two dimensions, the final post is only partially wet. ($X$ label). (d) The interface in the negative $x$-direction is disconnected, and pinned by the sides of the final post ($*$ label).}
 \label{fig:modes}
\end{figure}
\begin{figure}
\centering
\includegraphics[width=180mm]{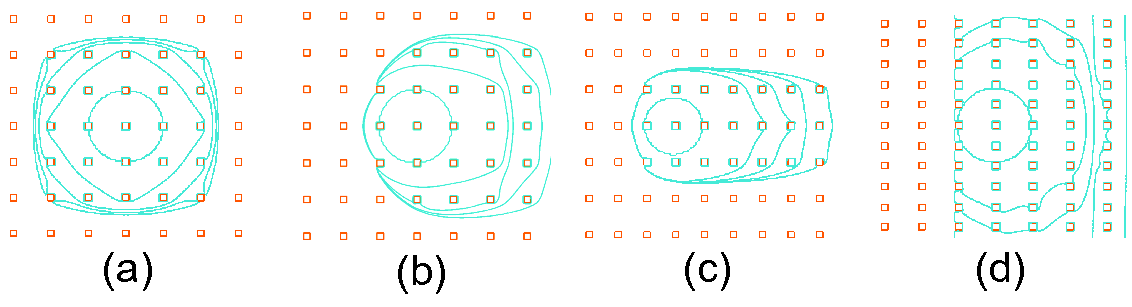}
\caption{Three dimensional spreading for $\theta = 45^o$ and (a) $\phi= 90^o$ , $a_x=a_y=4.6w$. (b) $\phi= 60^o$ , $a_x=a_y=4.6w$. (c) $\phi= 60^o$ , $a_x=3.6w$, $a_y=4.6w$. (d) $\phi= 60^o$ , $a_x=4.6w$, $a_y=2.53w$. Contours showing the interface position at the base of the drop are taken every $4\cdot 10^4$ time steps.}
 \label{fig:3D}
\end{figure}

Fig.~\ref{fig:1D} shows snapshots of the imbibition process as a function of time for the quasi-2D geometry and a contact angle $\theta=45^o$. The interface advances in the positive $x$-direction, but not in the negative $x$-direction, because of pinning on the posts. As pointed out in Chu et al. \cite{chu2010uni} a good understanding of why this occurs follows from assuming that the interface is pinned at the top corner of the posts and ignoring any interface curvature along $y$. We illustrate this situation in Fig.~\ref{fig:layout}b. The bottom of the interface will advance along the surface until it reaches the equilibrium contact angle $\theta$. If this enables the interface to reach the next post it will wet this post and move forwards, if not, it will remained pinned because any forward motion will increase the free energy. Because of the two dimensional nature of the model, it is easy to work out the threshold Young angles for spreading in the two directions as a function of the post geometry. Defining these as $\theta^+$ and $\theta^-$ for spreading along $+x$ and $-x$ respectively,  gives \citep{chu2010uni}
\begin{equation}
\label{eq:crit+}
\theta_{cr,\pm X} = \tan^{-1} \left(\frac{H}{a_x-w \mp H\cot \left( \phi \right)} \right).
\end{equation}
Thus there are three regimes: the interface can remain pinned in both directions, advance just along $+x$, or move forward in both directions. The different regimes predicted by Eq.~(\ref{eq:crit+}) are indicated in Fig.~\ref{fig:spreading} as a function of the Young angle $\theta$ and the post inclination $\phi$.  Note that, for $\phi=90^0$, $\theta^+=\theta^-$ as expected. The value of $a_x$  also affects the transition between different spreading regimes, as is apparent from Eq.(7): closer posts will ease the spreading, while posts that are further apart will make it more difficult. Our choice of $a_x=4.6w$ allows us to observe the different spreading regimes over the range of contact angles $\theta$ and slanting angles $\phi$ we consider. Fig.~\ref{fig:spreading} also shows the results of simulations for the quasi-2D geometry.\\

All three regimes are reproduced in the simulations. The analytic model gives a good account of the boundaries between them, but spreading in both directions is slightly more difficult than predicted by the 2D theory. The simulations allow us to identify this as being due to the details of the interface pinning on the posts. For an interface advancing along the positive $x$-axis, where the post points towards the direction of travel, we observe three different possible pinning mechanisms, labelled $||$, $B$ and $X$ in Fig.~\ref{fig:modes}. If the post is almost vertical, the leading interface is disconnected and is pinned to the vertical sides of the post ($||$ label). For a more pronounced post tilt the interface remains disconnected, but does not reach the top of the final post ($B$ label). For large tilt the leading interface is connected and has reached the equilibrium contact angle on the substrate. However, in contrast to two dimensions, the final post is only partially wet. ($X$ label). The situations, $B$ and $X$, where the interface has only reached the top of the penultimate post is only observed for slanted posts. It occurs because the interface can take the correct contact angle on the final post without a large penalty in curvature energy. The interface configuration resembles that in the partially suspended state identified in Kusumaatmaja and Yeomans \cite{kusumaatmaja2009anisotropic}.\\

 We considered a small spacing between posts to facilitate spreading over a wide range of wetting angles. It is worth noticing however that, if the spacing between posts were increased in the spreading direction, configurations analogous to the 2D theory would likely appear, with pinning on the final rather than penultimate row of posts.\\

In the hard direction for spreading, $-x$, the interface is pinned at the edges of the final line of posts, adjusting to their slope, as shown in Fig.~\ref{fig:modes}, label $*$. Bidirectional spreading only occurs for very low contact angles $\theta <  30^o$ or posts close to vertical $\phi > 70^o$.\\

These simulations correspond to  quasi-static spreading, with the fluid reservoir replenished very slowly. Borderline configurations between different wetting modes (such as $\phi=70^o$, $\theta=60^o$) are very sensitive to exact details of the position and filling speed of the reservoir. This is expected because the free energy barriers and capillary forces driving the flow are very small. A comparison between  two resolutions used shows that, as expected, spreading is slightly more difficult for a narrower interface.\\

We also note that, if the rate at which fluid is added to the reservoir is increased, the resulting inertia allows the fluid to de-pin from the top of the posts, forming a spherical cap.\\

We next report a full three-dimensional simulation which allows the fluid to spread along both $x$ and $y$. The reservoir is defined as a circular region of radius $10w$ in the centre of the domain, and the contact angle is $\theta = 45^o$.  In Fig.~\ref{fig:3D} we plot contour lines showing the spreading of the drop base for subsequent time steps. The first plot is for vertical posts; as expected it reflects the symmetry of the lattice. In the second plot the lattice spacing is the same, but the posts are now slanted with  $\phi= 60^o$. In the slanting direction the behaviour is consistent with the quasi-2D model, with the liquid spreading only in the positive $x$ direction. The spreading in the $y$ direction is comparable to the vertical post case.  The overall dynamics closely resembles the imbibition observed in experiments by Chu et al.\cite{chu2010uni}.\\

In Fig.~\ref{fig:3D}c the posts are closer in the x-direction. One can see that the asymmetry of the spreading becomes more pronounced, with the fluid spreading easily from one row to the next along x, while only spreading very slowly along y. This occurs because the thermodynamic driving force for spreading is much stronger in the x direction. For a contact angle of $45^o$  there is slow spreading along y, however for a larger contact angle the fluid will remained pinned along y  (the crossover can be estimated from Eq.\ref{eq:crit+}, taking $\phi =90^o$ in Figure 4), and/or x (as seen in the quasi-2D simulations reported in Fig.4).\\

 Eventually, in Fig.~\ref{fig:3D}d, we simulate densely packed posts in the $y$ direction, while keeping the same $x$ spacing as in Fig.~\ref{fig:3D}b. The behavior is now quite different, with the spreading happening first in the transverse direction, and only subsequently along the slanting direction. The unidirectionality is however maintained.\\

These results indicate that the final shape of a spreading drop can be tuned in detail by varying the lattice geometry and tilt angle of the posts, while retaining the relevant property of unidirectionality.
\section{Conclusions and outlook}

We have applied a lattice Boltzmann algorithm for two phase flow to model the spreading of liquid drops on a surface patterned by a lattice of slanted micro-posts. Gibb's pinning of the interface on the sides or top of the posts led to unidirectional spreading over a wide range of fluid-substrate contact angles and inclination angles of the posts. Regimes for spreading in no, one or two directions were identified, and shown to agree well with a two-dimensional theory proposed by Chu et al\cite{chu2010uni}. A more detailed numerical analysis of the contact line configurations enabled us to understand deviations from the two dimensional model, and to identify the configurations of the pinned interfaces.\\

The final drop shape depends on spacing of the post lattice, the contact angle, and the geometry and  inclination of the posts. Our simulations correspond to slow spreading, but inertial terms will also alter the final drop configuration. Thus there are many, varied possibilities to use slanted posts to control drop shapes or the direction of a flowing stream of fluid. Contact angles can be varied in situ by electrowetting, and it would be of interest to design substrates with addressable posts where the contact angle of each of the posts could be varied independently to allow steering of microfluidic flows.

\section*{Acknowledgements}
AC acknowledges the NanoVation consortium for funding this research. He also acknowledges Otto M\o nsted fonden, Reinholdt W. Jorck fonden and P.A. Fisker fonden for travel and accommodation support during the collaboration between the institutes. MLB acknowledges the support of the Portuguese Foundation for Science and Technology (FCT), through the grants SFRH/BDP/73028/2010 and PEst-OE/FIS/UI0618/2011. JMY acknowledges support from the ERC Advanced Grant MiCE.

\footnotesize{
\bibliography{Main} 
\bibliographystyle{plain} 

\end{document}

\bibliography{Main} 
\bibliographystyle{plain} 
